\documentstyle[12pt]{article}

\newtheorem{theorem}{Theorem}
\newtheorem{lemma}{Lemma}
\newtheorem{remark}{Remark}

\begin{document}

\def\prf{\noindent{\bf Proof.\ \,}}
\def\prfe{\hspace*{\fill} $\Box$}
\def\n#1{\vert #1 \vert}
\def\nn#1{\Vert #1 \Vert}
\def\lap{\bigtriangleup}
\def\be{\begin{equation}}
\def\ee{\end{equation}}
\def\bea{\begin{eqnarray}}
\def\eea{\end{eqnarray}}
\def\beas{\begin{eqnarray*}}
\def\eeas{\end{eqnarray*}}
 
\def\R{{\rm I\kern-.1567em R}}
\def\N{{\rm I\kern-.1567em N}}
 
\def\Tr{\mbox{\rm Tr}\,}

\def\C{{\cal C}}
\def\S{{\cal S}}
\def\G{{\cal G}}
\def\H{{\cal H}}

\title{Existence and nonlinear stability of steady states 
       of the Schr\"odinger-Poisson system}
\author{ Peter A.~Markowich${}^\ast$, Gerhard Rein${}^\dagger$, 
         Gershon Wolansky${}^\ddagger$\\
         ${}^\ast$ Mathematisches Institut der Universit\"at Wien \\
         Boltzmanngasse 9, 1090 Vienna, Austria\\
         ${}^\dagger$ International Erwin Schr\"odinger Institute\\ 
         Boltzmanngasse 9, 1090 Vienna, Austria\\       
         ${}^\ddagger$ Technion, 32000 Haifa, Israel
        }
\date{}
\maketitle

\begin{abstract}
We consider the Schr\"odinger-Poisson system in the 
attractive
(plasma physics) Coulomb case.
Given a steady state from a certain class
we prove its nonlinear stability, using an appropriately
defined energy-Casimir functional
as Lyapunov function. To obtain such steady states
we start with a given Casimir functional and
construct a new functional which is
in some sense dual to the corresponding energy-Casimir 
functional. This dual functional has a 
unique maximizer which
is a steady state
of the Schr\"odinger-Poisson system and lies in the 
stability class.
The steady states are parametrized
by the equation of state, giving the occupation 
probabilities of the 
quantum states as a strictly decreasing function of 
their energy levels.

\end{abstract}

{\bf Acknowledgement:} This research was supported 
by the OEAD, the International Erwin Schr\"odinger 
Institute in Vienna, the Wittgenstein 2000
prize of P.~A.~M.\ funded by the Austrian FWF, and 
the EU-funded TMR-network.

\section{Introduction}
\setcounter{equation}{0}

A large ensemble of  charged quantum particles
interacting only by the electrostatic field which 
they create collectively can be described by the
Schr\"odinger-Poisson system:
\be \label{schr}
i \frac{\partial\psi_k}{\partial t} = 
- \Delta\psi_k + V\psi_k,
\ee
\be \label{poi}
\lap V = -n,
\ee
\be \label{dens}
n=\sum_{k=1}^\infty \lambda_k |\psi_k|^2 .
\ee
Here $\psi_k = \psi_k(t,x)$ is the wave function of the 
$k$th state, 
$k \in \N$,
$\lambda_k\geq 0$ denote the corresponding occupation 
probabilities
normalized such that $\sum_k\lambda_k =1$,
$n=n(t,x)$ is the number density, and
$V=V(t,x)$ the self-consistent potential of the ensemble.
In order to avoid continuous spectra we shall analyze this
system on a bounded domain $\Omega\subset\R^3$ with 
sufficiently smooth
boundary, and we supplement it with Dirichlet boundary 
conditions: 
\be \label{bc}
\psi_k(t,x) = 0,\ V(t,x) = 0,\ t \geq 0,\ 
x \in \partial \Omega,\ k \in \N.
\ee
We could also consider the system on the whole space $\R^3$
and add to $V$ an appropriate exterior potential $V_e$. 
Initial data are given by a complete orthonormal system
$(\psi_k(\cdot,0))$ in $L^2(\Omega)$.
We refer to \cite{APT,BM,GIMS,Mar1} for background 
information on the Schr\"odinger-Poisson system 
(\ref{schr}), (\ref{poi}), (\ref{dens}).

In terms of the density operator $R(t)$
of the system, a time dependent, hermitian, positive 
trace-class operator
acting on the Hilbert space $L^2(\Omega)$, the time 
evolution is given by the von-Neumann-Heisenberg equation
\[
i\frac{\partial R}{\partial t} = [H_V,R].
\]
Here the Hamiltonian is defined as 
$H_V := - \lap + V(t,x)$
with potential $V$ given as the solution of the Poisson
equation (\ref{poi}) with Dirichlet boundary condition, 
and $n(t,x):=\rho(t,x,x)$
where $\rho(t,x,y)$ is the kernel of the operator $R(t)$. 
The Schr\"odinger-Poisson picture and the Heisenberg
picture are equivalent: 
Let $(\phi_k)$ be a complete orthonormal sequence
of eigenvectors of $R(0)$ with eigenvalues
$(\lambda_k)$ and let $(\psi_k(t,\cdot))$ be the solution of 
the Schr\"odinger-Poisson system (\ref{schr})--(\ref{bc})
with initial data $\psi_k(0)=\phi_k$. Then
$\rho(t,x,y)=\sum_k\lambda_k \psi_k(t,x)\, \bar\psi_k(t,y)$
defines the kernel of an operator $R(t)$ which solves the
von-Neumann-Heisenberg equation with the corresponding
initial datum, and vice versa.   

The Schr\"odinger-Poisson picture is more
suitable for our present purposes, which are as follows:
We investigate the nonlinear stability
of certain steady states of the Schr\"odinger-Poisson system,
i.~e., of solutions of the form 
$\psi_k (t,x)=e^{i \mu_k t}\phi_k(x)$
with energy levels $\mu_k \in \R$,
and we prove the existence
of such steady states. To our knowledge, the stability 
problem has not yet been investigated. 
The existence of steady states has been considered by 
different methods before, cf.\ \cite{Mar1,N1,N2,N3}.
 
Our approach is motivated by analogous results for the 
Vlasov-Poisson system which arises as the classical limit
of the Schr\"odinger-Poisson system. Both systems share 
the following property:
The total energy of the system is conserved along 
solutions---indeed,
the dynamics can be interpreted as the ``Hamiltonian flow''
induced by the energy functional---, but the steady states 
are not critical points of the energy. 
On the other hand, there exist additional
conserved quantities, the so-called Casimir 
functionals \cite{Cas}, 
such that a given steady state is a critical 
point for the appropriately 
chosen energy-plus-Casimir functional $\H_C$.
The energy-Casimir method was first used to prove
genuine, nonlinear stability for fluid-flow problems by
{\sc Arnol'd} in the 1960's, cf.\ \cite{Ar1,Ar2}. Some 
of the background of this method can be found in \cite{HMRW}.
More recently, the energy-Casimir method 
was adapted to problems in kinetic theory,
in particular the Vlasov-Poisson system, 
cf.\ \cite{G1,G2,GR1,GR2,GR3,R1,R2,Wo1,Wo2}. 
When applying this method
there is a sharp contrast between the plasma 
physics situation and 
the stellar dynamics one, where the sign in the 
Poisson equation is reversed:
The quadratic part in the expansion of the 
energy-Casimir functional
at the steady state is positive definite in the plasma
physics case while it is indefinite in the stellar dynamics 
case. Therefore, in the former case the method applies 
in a straight forward manner, cf.\ \cite{R1}, 
while in the latter case 
a careful investigation of the behavior of the energy-Casimir
functional along minimizing sequences is needed and leads
to nonlinear stability only for such steady states which are
obtained as minimizers of this functional.
The present paper addresses the plasma-physics case for 
the Schr\"odinger-Poisson system, and thus the approach 
should be 
more like the former case for the Vlasov-Poisson system.

This is indeed so: In Section~3 we show that
steady states $(\psi_0,\lambda_0)$
from a specified class are nonlinearly stable,
and we do so by estimating 
$\H_C (\psi,\lambda) - \H_C(\psi_0,\lambda_0)$ 
from below by an expression which is quadratic
in $(\psi,\lambda)-(\psi_0,\lambda_0)$, where
$(\psi,\lambda)$ is some other, `close-by' state. 
In Section~4 we construct a functional which is
in some sense dual to a given
energy-Casimir functional. As shown in Section~5
this dual functional has a unique
maximizer, which is a steady state, and nonlinearly stable 
by Section 3. 
We emphasize that---as opposed to
the stellar-dynamics situation for the Vlasov-Poisson 
system---the stability analysis and the existence analysis 
are independent from each other; the connecting Section~4 
is included to put both parts into a common perspective. 
Before going into all this we introduce the class of 
steady states respectively Casimir functionals under 
consideration, derive some
preliminary estimates, and fix some notation.

\section{Preliminaries}
\setcounter{equation}{0}

As state space for the 
Schr\"odinger-Poisson system we use the set
\beas
\S := \Bigl\{ (\psi,\lambda)
&|&
\psi = (\psi_k)_{k \in \N} \subset 
H^1_0(\Omega) \cap H^2(\Omega)\\
&&
\mbox{is a complete orthonormal system in}\ L^2(\Omega),\\
&&
\lambda = (\lambda_k)_{k \in \N} \in l^1 
\ \mbox{with}\ \lambda_k \geq 0,\ k \in \N,\\
&&
{\sum}_k \lambda_k \int \n{\lap \psi_k}^2 < \infty \Bigr\};
\eeas
$\sum_k$ always means $\sum_{k=1}^\infty$.
Our notation for the Sobolev spaces $H^2$ and $H^1_0$
is the standard one; by $\nn{\cdot}_p$ we will denote the 
norm in the usual $L^p$ space.
For $(\psi,\lambda) \in \S$ we have
\[
n_{\psi,\lambda}:= {\sum}_k \lambda_k \n{\psi_k}^2 
\in L^2(\Omega),
\]
and $V_{\psi,\lambda}$  denotes
the Coulomb potential induced by $n_{\psi,\lambda}$, i.~e.,
\[
\lap V_{\psi,\lambda} = - n_{\psi,\lambda}\ 
\mbox{on}\ \Omega,\
V_{\psi,\lambda} = 0\ \mbox{on}\ \partial \Omega;
\]
note that 
$V_{\psi,\lambda}\in H^1_0(\Omega)\cap H^2(\Omega)$
by the energy bound and Sobolev inequalities.
For every initial state 
$(\psi(0),\lambda) \in \S$
there is a unique strong solution 
$[0,\infty[ \ni t \mapsto \psi(t)$ 
of (\ref{schr})--(\ref{bc})
with $(\psi(t),\lambda)\in \S$, 
cf.~\cite{BM}. 
Throughout the paper, potentials $V$ are real-valued while 
quantum states $\psi_k$ are complex-valued.
The energy of a state $(\psi,\lambda) \in \S$ is defined as
\beas
\H (\psi,\lambda) 
&:=&
{\sum}_k \lambda_k \int|\nabla\psi_k|^2 +
\frac{1}{2}
\int n_{\psi,\lambda} V_{\psi,\lambda} \\
&=&
{\sum}_k \lambda_k \int|\nabla\psi_k|^2 +
\frac{1}{2}
\int |\nabla V_{\psi,\lambda}|^2;
\eeas
integrals always extend over the set $\Omega$.
The energy is conserved along
solutions of the Schr\"odinger-Poisson system , 
indeed, the system (\ref{schr})--(\ref{bc})
can be written in the form
\beas
i \frac{\partial\psi_k}{\partial t} 
&=& 
-\frac{1}{2\lambda_k} \delta_{\bar\psi_k} \H,\\
i \frac{\partial\bar\psi_k}{\partial t} 
&=& 
-\frac{1}{2\lambda_k} \delta_{\psi_k}\H,\\
\frac{d \lambda_k}{dt} 
&=& 
0,
\eeas
where the bar denotes complex conjugation.

To assess the stability of a given steady state
we employ an energy-Casimir functional
\[
\H_C (\psi,\lambda) := {\sum}_k C(\lambda_k) + 
\H(\psi,\lambda)
\]
with the real-valued function $C$ defined appropriately.
Clearly, $\H_C$ is a conserved
quantity for the Schr\"odinger-Poisson system.

The class of functions which generate the Casimir
functionals will now be specified: 
We say that a function 
$f:\R \to \R$ 
{\em is of Casimir class $\C$} iff it has the following 
properties:
\begin{itemize}
\item[(i)]
$f$ is continuous with
$f(s) > 0,\ s \leq s_0$ and $f(s)=0,\ s \geq s_0$ for some
$s_0 \in ]0,\infty]$,
\item[(ii)]
$f$ is strictly decreasing on $]-\infty,s_0]$ with
$\lim_{s\to -\infty} f(s)=\infty$,
\item[(iii)]
there exist constants $\epsilon > 0$ and $C >0$ such that
\[
f(s) \leq C(1+s)^{-7/2 - \epsilon},\ s \geq 0.
\]
\end{itemize}
For $f \in \C$,
\be \label{fdef}
F(s) := \int_s^\infty f(\sigma)\, d\sigma,\ s \in \R,
\ee
defines a decreasing,
continuously differentiable, and non-negative function
which is strictly convex on its support, 
and
\[
F(s) \leq C(1+s)^{-5/2 - \epsilon},\ s \geq 0.
\]
In passing we note that by adjusting various exponents our
results easily extend to general space dimensions.

\smallskip

\begin{remark}{\rm 
\begin{itemize}
\item[(a)]
A typical example for $f\in\C$ is the
Boltzmann distribution $f(s)=e^{-\beta s}$ with $\beta > 0$,
where the cut-off level $s_0 =\infty$.
Another example, which also decays exponentially for 
$s \to \infty$,
is given by the Fermi-Dirac statistics:
\[
f(s):= C \int_{\R^3} 
\frac{dv}{\epsilon + e^{|v|^2/2 +s}},\ s \in \R,
\]
where $C>0$ and $\epsilon > 0$ are positive parameters. 

A function $f$ with $f(s)=0$ for $s>s_0$ with $s_0 \in \R$ 
will yield a steady state consisting of a finite number of 
quantum oscillators.
\item[(b)]
We could generalize the assumption (iii) to requiring that 
both $f(-\Delta +V)$ and $F(-\Delta+V )$
are of trace class for (smooth) potentials $V\geq 0$, 
cf.\ Lemma~\ref{traceclass} (b) below. 
However, we prefer to keep our assumptions on $f$ explicit.
\end{itemize} }
\end{remark}

\begin{lemma} \label{traceclass}
Let $f \in \C$. 
\begin{itemize}
\item[{\rm (a)}]
For every $\beta > 1$ there exists $C=C(\beta)\in \R$ such 
that
\[
F(s) \geq - \beta s + C,\ s \leq 0.
\]
\item[{\rm (b)}]
Let $V \in H^1_0(\Omega)$
be non-negative on $\Omega$. Then both
$f(-\Delta +V)$ and $F(-\Delta+V)$
are trace class.
\end{itemize}
\end{lemma}

\noindent
\prf
Part (a) is straight forward from assumption (ii) and the 
definition of $F$.
As to (b), let $(\mu_k)$ denote the sequence of eigenvalues
of $-\lap + V$.
Then, since $V$ is non-negative and $F$
decreasing,
\[
{\sum}_k F(\mu_k) \leq {\sum}_k F(\mu_k^0)
\]
where $\mu_k^0$ denote the eigenvalues of $-\lap$.
For the latter we have the well-known estimate
that the number of such eigenvalues less than some 
$\mu \in \R$
grows like $\mu^{3/2}$ for $\mu \to \infty$,
which implies that the right hand sum is
finite, and $F(-\lap +V)$ is trace class.
Since $f$ decays faster than $F$ the same holds
true for $f(-\lap +V)$. \prfe

At several points the following technical observation will
be useful:

\begin{lemma} \label{specconv}
For $\psi \in H^1_0(\Omega) \cap H^2(\Omega)$ with 
$\nn{\psi}_2 =1$
and $V \in H^1_0(\Omega),\ V\geq 0,$ we have
\[
F\left(\langle \psi, (-\lap + V) \psi \rangle\right) \leq 
\langle \psi, F(-\lap + V) \psi \rangle
\]
with equality if $\psi$ is an eigenstate of $-\lap + V$.
\end{lemma}

\prf
Denoting the spectral measure associated with $-\lap +V$
and $\psi$ by $d\sigma$ the claim translates into the 
inequality
\[
F\left(\int \sigma d\sigma \right)
\leq \int F(\sigma) d\sigma 
\]
which holds due to the convexity of $F$ and Jensen's 
inequality.
\prfe

To conclude this section we make precise the notion of a 
steady
state of the Schr\"odinger-Poisson system:
A quadruple $(\psi_0,\lambda_0,\mu_0,V_0)$ with 
$(\psi_0,\lambda_0) \in \S$, 
$\mu_0 = (\mu_{0,k}) \in \R^\N$, and 
$V_0 \in H^2(\Omega) \cap H^1_0(\Omega)$ is a 
{\em steady state of the Schr\"odinger-Poisson system
(\ref{schr})--(\ref{bc})}\, iff
\be \label{stschr}
(-\lap +V_0)\, \psi_{0,k} = \mu_{0,k}\, 
\psi_{0,k},\ k \in \N,
\ee
and
\be \label{stpoisson}
\lap V_0 = - n_0 = - {\sum}_k \lambda_{0,k} |\psi_{0,k}|^2 ,
\ee
where the energy levels $\mu_{0,k}$
and occupation probabilities $\lambda_{0,k}$ are
related through an equation of state of the form
\be \label{eqstate}
\lambda_{0,k} = f(\mu_{0,k}),\ k \in \N,
\ee
with some $f \in \C$. 

\begin{remark}\label{ststins}{\rm
If $(\psi_0,\lambda_0,\mu_0,V_0)$ satisfies the equations
(\ref{stschr}), (\ref{stpoisson}), (\ref{eqstate}) with
$f \in \C$ then the estimate
\[
{\sum}_k \lambda_{0,k} \nn{\psi_{0,k}}_{H^2}^2 < \infty .
\]
follows and thus in particular $(\psi,\lambda)\in \S$.
To see this we use (\ref{stschr}) and estimate
\[ 
{\sum}_k \lambda_{0,k} \nn{\nabla \psi_{0,k}}_2^2 + 
\int|\nabla V_0|^2 
=
{\sum}_k \mu_{0,k} f(\mu_{0,k})
\leq
C {\sum}_k (1+\mu_{0,k})^{-(5/2 + \epsilon)} < \infty 
\]
by assumption (iii) on $f$ and the asymptotic behaviour
of $\mu_{0,k}$. Thus, by the Sobolev inequality,
\[
\nn{n_0}_3 \leq {\sum}_k \lambda_{0,k} \nn{\psi_{0,k}}_6^2 
< \infty,
\]
and $V_0 \in W^{2,3}(\Omega) \subset L^\infty (\Omega)$ 
follows. Again from (\ref{stschr})
we conclude that
\beas
{\sum}_k \lambda_{0,k} \nn{\lap \psi_{0,k}}_2^2
&\leq&
C \left({\sum}_k \lambda_{0,k} \mu_{0,k}^2 
+ {\sum}_k \lambda_{0,k} \right)\\
&\leq&
C \left(1+{\sum}_k (1+\mu_{0,k})^{-(3/2+\epsilon)}\right) < 
\infty.
\eeas
}
\end{remark}

Given $f\in \C$ we still need to
specify the corresponding Casimir functional: With 
$F$ given by (\ref{fdef}), 
its Legendre or Fenchel transform is defined by
\be \label{fstardef}
F^\ast (s):= \sup_{\lambda \in \R} 
(\lambda s - F(\lambda)),\ s \in \R,
\ee
and the energy-Casimir functional corresponding to $f$ is 
\be \label{encasdef}
\H_C (\psi,\lambda) := {\sum}_k F^\ast (-\lambda_k) + 
\H(\psi,\lambda),\
(\psi,\lambda) \in \S.
\ee
Note that since $F'=-f$ has an inverse on $]-\infty,s_0[$,
\be \label{fstardefint}
F^\ast (s) = \int_{-s}^0 f^{-1} (\sigma)\, d\sigma
\ee
for $-\infty = - f(-\infty) < s \leq 0$, and all $-\lambda_k$
lie in this interval.

Obviously, only the values of $f \in \C$ on the interval
$]0,\infty[$ are significant for the following theory.
However, for technical reasons we consider the functions 
$f$ defining the
equations of state as defined on all of $\R$.

\section{Stability}
\setcounter{equation}{0}

In the present section we shall establish
the following result on nonlinear stability:
\begin{theorem} \label{stability}
Let $(\psi_0,\lambda_0,\mu_0,V_0)$ be a steady
state of the Schr\"odinger-Poisson system with
\[
\lambda_{0,k} = f(\mu_{0,k}),\ k \in \N,
\]
for some $f \in \C$, and $(\psi_0,\lambda_0)\in \S$. 
Then this steady state is nonlinearly stable in the 
following sense: 
If $t \mapsto (\psi(t),\lambda)$ is a solution of the 
Schr\"odinger-Poisson system with initial datum
$(\psi(0),\lambda) \in \S$ then
\[
\frac{1}{2}
\left|\left| \nabla V_{\psi(t),\lambda} - 
\nabla V_0\right|\right|_2^2 
\leq \H_C (\psi(0),\lambda) - \H_C (\psi_0,\lambda_0),
\ t \geq 0.
\]
\end{theorem}
We recall that $\H_C$ is defined by (\ref{encasdef}) 
for the given function $f$ and note that, clearly, 
the right hand side in the estimate above
becomes arbitrarily small if $(\psi(0),\lambda)$
is close to $(\psi_0,\lambda_0)$ in the appropriate 
topology.
The main step in the proof of Theorem~\ref{stability} 
is to show the following estimate:
\begin{lemma} \label{stablemma}
Let $V \in H^1_0(\Omega),\ V \geq 0$.
Then 
\[
{\sum}_k 
\left[F^\ast(-\lambda_k)
+ \lambda_k \int\left[|\nabla\psi_k|^2 + 
V |\psi_k|^2\right] \right]
\geq 
- \Tr [F(-\lap + V)],\ (\psi,\lambda) \in \S,
\]
with equality for $(\psi,\lambda)= (\psi_V,\lambda_V)$, where
$\psi_V = (\psi_{V,k})\in H^1_0(\Omega)^\N$ is an orthonormal
sequence of eigenfunctions of $-\lap +V$ with eigenvalues
$\mu_V = (\mu_{V,k})$, and 
$\lambda_V = (\lambda_{V,k})=(f(\mu_{V,k}))$.
\end{lemma}

\prf
The fact that $F$ and $F^\ast$ are related by conjugacy 
implies that
\bea
F^\ast (-\lambda) + \lambda \mu
&\geq&
\inf_{s \in \R}[F^\ast (-s) + s \mu] \nonumber\\
&=&
- \sup_{s \in \R}[- F^\ast (s) + s \mu] = - F^{\ast\ast}(\mu)
\nonumber \\
&=&
- F(\mu),\ \lambda, \mu \in \R. \label{conjug}
\eea
We substitute $\lambda_k$ for $\lambda$ and
\[
\mu_k := \int\left[|\nabla\psi_k|^2 + V |\psi_k|^2\right]
= \langle \psi_k, (-\lap + V) \psi_k \rangle
\]
for $\mu$ and sum over $k$ to find
\beas
{\sum}_k
\left[F^\ast(-\lambda_k)
+ \lambda_k \int\left[|\nabla\psi_k|^2 + 
V |\psi_k|^2\right] \right]
&\geq& 
- {\sum}_k F(\langle \psi_k, (-\lap + V) \psi_k \rangle) \\
&\geq&
- {\sum}_k \langle \psi_k, F(-\lap + V) \psi_k \rangle) \\
&=&
- \Tr [F(-\lap + V)]
\eeas
by Lemma~\ref{specconv} and the definition of trace. 

Now suppose that $(\psi,\lambda)= (\psi_V,\lambda_V)$. Since
by definition each $\psi_{V,k}$ is an eigenfunction
of $-\lap +V$ the 
$\mu_k$ defined above are the corresponding eigenvalues 
$\mu_{V,k}$,
and
\[
\Tr [F(-\lap + V)] = {\sum}_k F(\mu_{V,k}) .
\]
On the other hand we have 
$\lambda_{V,k}=f(\mu_{V,k}) = - F'(\mu_{V,k})$
which by conjugacy
is equivalent to $\mu_{V,k}= {F^\ast}'(-\lambda_{V,k})$, 
$k \in \N$.
This implies that
\[
{\sum}_k F(\mu_{V,k}) 
= - 
{\sum}_k \left[F^\ast(-\lambda_{V,k}) + 
\lambda_{V,k}\, \mu_{V,k} \right],
\]
and the proof is complete. \prfe

\begin{remark}\label{traceeq}
{\rm
In Lemma~\ref{stablemma} equality holds if {\em and only if}
$(\psi,\lambda)= (\psi_V,\lambda_V)$. This follows from the 
strict convexity of $F$, but we make no use of this 
observation in the rest of the paper.
}
\end{remark}

\noindent
{\bf Proof of Theorem~\ref{stability}.\ \,}
Let $V=V_{\psi,\lambda}$ be the potential
induced by $(\psi, \lambda) \in \S$. Then
\beas
&&
\frac{1}{2}
\left|\left| \nabla V - \nabla V_0\right|\right|_2^2\\
&&=
\frac{1}{2} \int |\nabla V|^2 + \int \lap V \,V_0 + 
\frac{1}{2} \int |\nabla V_0|^2\\
&&=
\H_C(\psi,\lambda) -
\left[ {\sum}_k\left(F^\ast (-\lambda_k) + 
\lambda_k \int |\nabla \psi_k|^2\right)- 
\frac{1}{2} \int |\nabla V_0|^2
- \int \lap V \,V_0\right] \\
&&=
\H_C(\psi,\lambda) -
\left[ {\sum}_k\left(F^\ast (-\lambda_k) + 
\lambda_k \int \left[|\nabla \psi_k|^2 +
V_0 |\psi_k|^2\right]\right)
- \frac{1}{2} \int |\nabla V_0|^2\right] \\
&&\leq
\H_C(\psi,\lambda) - 
\left[ - \Tr [F(-\lap + V_0)] - 
\frac{1}{2} \int |\nabla V_0|^2\right] \\
&&=
\H_C(\psi,\lambda) -
\left[ {\sum}_k\left(F^\ast (-\lambda_{0,k}) + 
\lambda_{0,k} \int (|\nabla \psi_{0,k}|^2 +
V_0 |\psi_{0,k}|^2)\right)
- \frac{1}{2} \int |\nabla V_0|^2\right] \\ 
&&=
\H_C(\psi,\lambda) - \H_C(\psi_0,\lambda_0),
\eeas
where we have used Lemma~\ref{stablemma} twice.
Given a solution with $(\psi (0),\lambda)\in \S$ we
may substitute $(\psi (t),\lambda)\in \S$ into this estimate,
and since $\H_C$ is constant along solutions
the assertion follows. \prfe
\section{Dual functionals}
\setcounter{equation}{0}

Our aim for the rest of this paper is to
prove the existence of steady states which satisfy the
assumption of our stability result. For each $f \in \C$
a corresponding steady state will be obtained as the unique
maximizer of an appropriately defined functional.
In the present section we derive this dual functional
from the energy-Casimir functional used in the stability
analysis. The relation between these functionals
is of interest in itself, but it is not used
in the proofs of our results. Throughout this
section we fix an element $f \in \C$.
We move to the dual functional in two steps. First we 
apply the saddle point principle and define,
for $\Lambda > 0$ fixed,
\beas
\G (\psi, \lambda, V, \sigma) 
&:=& 
{\sum}_k \left[F^\ast(-\lambda_k)
+ \lambda_k \int\left[ |\nabla\psi_k|^2 + 
V|\psi_k|^2\right]\right]
- \frac{1}{2}\int |\nabla V|^2 \\
&&
{}+ \sigma \, \left[{\sum}_k\lambda_k-\Lambda \right] 
\eeas
where $\psi =(\psi_k)$ is again an orthonormal system in 
$L^2(\Omega)$,
$\lambda \in l_+^1 =\{ (\sigma_k) \in l^1 | 
\sigma_k \geq 0,\ k \in \N\}$, 
and $V \in H^1_0(\Omega)$
may now vary independently of $\psi$ and $\lambda$.
The role of the parameter
$\sigma \in \R$ (Lagrange multiplier)
will become clear shortly; the relation between $\H_C$
and this new functional is as follows:

\begin{remark}
{\rm
For any $\psi, \lambda, \sigma$,
\be \label{grelhc}
\sup_{V} \, \G (\psi, \lambda, V, \sigma)
= \H_C(\psi,\lambda) + \sigma\,
\left[{\sum}_k\lambda_k -\Lambda\right],
\ee
and the supremum is attained at $V=V_{\psi,\lambda}$.
In fact, integration by parts and some computations show that
\[
\G (\psi, \lambda, V, \sigma) =
\H_C(\psi,\lambda) + \sigma\, 
\left[{\sum}_k\lambda_k -\Lambda\right]
-\frac{1}{2} \nn{\nabla V_{\psi,\lambda} - \nabla V}_2^2 .
\]
}
\end{remark}
As second step on our way to a dual variational formulation
we reduce the functional $\G$ to a functional of $V$ and 
$\sigma$ as follows:
\be \label{phidef}
\Phi(V,\sigma) := \inf_{\psi,\lambda} 
\G (\psi, \lambda, V, \sigma)
\ee
where the infimum is taken over all $\lambda \in l_+^1$ and 
all orthonormal sequences $\psi$ in $L^2 (\Omega)$.
It is this functional which will have a unique maximizer
in the next section, which is then a steady state.
First however, we need to bring it into a different form: 

\begin{remark}
{\rm
The infimum in the definition of $\Phi$ is attained
at $\psi=(\psi_{V,k})$, an orthonormal sequence of
eigenstates of $-\lap + V$ with corresponding
eigenvalues $\mu_{V,k}$, and $\lambda=\lambda_V$
where $\lambda_{V,k}=f(\mu_{V,k} + \sigma),\ k \in \N$. 
Moreover,
\[
\Phi(V,\sigma)=-\frac{1}{2}\int|\nabla V|^2 -
\Tr\left[ F(-\Delta + V + \sigma)\right]- \sigma \, \Lambda .
\]
To see this, recall Lemma~\ref{stablemma} and 
Remark~\ref{traceeq}
and observe that $f(\cdot + \sigma) \in \C$ for any 
$\sigma \in \R$, provided $f \in \C$.
}
\end{remark}

\section{Existence of steady states}
\setcounter{equation}{0}

In the present section we shall for each state relation 
$f \in \C$ 
and each total charge $\Lambda > 0$ construct
a unique maximizer of the functional $\Phi$, which is then
a steady state of the Schr\"odinger-Poisson system.
We consider only non-negative potentials and use
the notation
\[
H^1_{0,+}(\Omega) := \{ V \in H^1_0 (\Omega) | V \geq 0 \}.
\]

\begin{theorem} \label{exstst}
Let $f \in \C$ and $\Lambda > 0$ be given.
The functional 
\[
\Phi : H^1_{0,+}(\Omega) \times \R \ni 
(V,\sigma) \mapsto 
-\frac{1}{2}\int|\nabla V|^2 -
\Tr\left[ F(-\Delta + V + \sigma)\right]- \sigma \, \Lambda
\]
is continuous, strictly concave, bounded
from above,  and coercive.
In particular, there exists a unique
maximizer $(V_0, \sigma_0)$ of $\Phi$.
If we define $\psi_0=(\psi_{0,k})$ as the orthonormal
sequence of eigenstates of the operator $-\lap +V_0$
with corresponding eigenvalues $\mu_{0,k}$ and 
$\lambda_{0,k}:= f(\mu_{0,k}+\sigma_0)$,
then $(\psi_0,\lambda_0,\mu_0,V_0)$
is a steady state of the Schr\"odinger-Poisson system
with $\sum_k \lambda_{0,k} = \Lambda$ and 
$(\psi_0,\lambda_0)\in \S$.
\end{theorem}
Note that $\sigma_0$ plays the role of a (constant) 
Fermi level here.

\noindent\prf
{\em $\Phi$ is strictly concave:}
The first term of $\Phi$ is evidently concave.
To show the strict concavity of the second term 
i.e., the strict convexity
of $\Tr\left[ F(-\Delta + V + \sigma)\right]$, 
let $(V_j,\sigma_j)\in H^1_{0,+} \times \R$, $j=1,2$,
$\alpha \in ]0,1[$, and 
$\phi \in H^2\cap H^1_0$. By convexity of $F$ and 
Lemma~\ref{specconv},
\beas
&&
F(\langle \phi,\alpha (-\lap + V_1 + \sigma_1)\phi +
(1-\alpha) (-\lap + V_2 + \sigma_2)\phi \rangle)\\
&& \qquad \qquad \leq
\alpha \langle \phi, 
F (-\lap + V_1 + \sigma_1)\phi \rangle +
(1-\alpha) \langle \phi, 
F(-\lap + V_2 + \sigma_2)\phi \rangle .
\eeas
Now we substitute $\psi_k$ for $\phi$, $(\psi_k)$ an 
orthonormal sequence of eigenstates of 
$\alpha (-\lap + V_1 + \sigma_1) + (1-\alpha) 
(-\lap + V_2 + \sigma_2)$,
and sum over $k$ to obtain the convexity estimate for
$\Tr\left[ F(-\Delta + V + \sigma\right]$.
If we have equality in this estimate then
\[
\langle \psi_k, F (-\lap + V_1 + \sigma_1)\psi_k \rangle
=
\langle \psi_k, F (-\lap + V_2 + \sigma_2)\psi_k \rangle,
\ k \in \N
\]
and thus $V_1=V_2$ and $\sigma_1=\sigma_2$.

\noindent 
{\em $\Phi$ is bounded from above and coercive:}
Since $F$ is non-negative, the critical case in the 
coercivity estimate is $\sigma <0$.
Let $\underline{\mu}_V$ denote the ground state energy 
of $-\Delta + V$ with corresponding ground state 
$\underline{\psi}_V$. 
Since $F$ is non-negative and satisfies estimate (a)
in Lemma~\ref{traceclass}
we have for $\sigma \leq -\underline{\mu}_V$,
\beas
\Phi(V,\sigma)
&\leq&
- \frac{1}{2}\int |\nabla V|^2 - 
\langle \underline{\psi}_V,F(-\lap +V+\sigma)
\underline{\psi}_V\rangle -\sigma \, \Lambda \\
&=&
- \frac{1}{2}\int |\nabla V|^2 - 
F(-\underline{\mu}_V+\sigma) -\sigma \, \Lambda \\
&\leq&
- \frac{1}{2}\int |\nabla V|^2 + (\beta - \Lambda)\,\sigma + 
\beta \underline{\mu}_V -C ,
\eeas
where we choose $\beta > \Lambda$.
Also 
\[
\underline{\mu}_V 
= 
\inf_{\phi \in H^1_0,\ \nn{\phi}_2=1} 
\int \left[-|\nabla \phi|^2 + V |\phi|^2\right]
\leq
\frac{1}{{\rm vol}\, \Omega}\int V \leq 
C_1 \nn{V}_{H^1_0},
\]
choosing $\phi := 1/\sqrt{{\rm vol}\, \Omega}$.  
Together with the estimate above and Poincar\'{e}'s 
inequality this implies that for 
$\sigma \leq - C_1 \nn{V}_{H^1_0}$ we have
\be \label{coerc1}
\Phi(V,\sigma) 
\leq
- C_2 \nn{V}^2_{H^1_0} + C_3 \nn{V}_{H^1_0} +
(\beta -\Lambda)\, \sigma + C_4
\ee
where the constants $C_1,C_2,C_3,C_4$ are positive and 
$\beta >\Lambda$,
cf.\ Lemma~\ref{traceclass} (a).
On the other hand, by the non-negativity 
of $F$ and Poincar\'{e}'s inequality,
\be \label{coerc2}
\Phi(V,\sigma) 
\leq
- C_2 \nn{V}^2_{H^1_0} - \sigma \, \Lambda,
\ee
and (\ref{coerc1}) and (\ref{coerc2}) together imply that
$\Phi$ is bounded from above and coercive.

\noindent
{\em Existence of a unique maximizer:}
The existence of a unique maximizer of $\Phi$
is standard, cf.\ for example \cite[Ch.~II, Prop.~1.2]{Ek},
provided $\Phi$ is upper semi-continuous. This in turn 
follows from the fact that $\Phi$ is concave and bounded 
from below, at least locally,
cf.\ \cite[Ch.~I, Lemma~2.1]{Ek}: The only term for which
this may not be immediately obvious is the trace term, but
\[
\Tr[F(-\lap + V + \sigma)]
\leq
{\sum}_k F(\mu_k + \sigma_0)
< \infty
\]
where $\mu_k$ are the eigenvalues of $-\lap +V$
and $\sigma \geq \sigma_0$ for
arbitrary $\sigma_0 \in \R$.

\noindent
{\em $(\psi_0,\lambda_0,\mu_0,V_0)$ is a steady state:}
Since $F'=-f$, the stationarity of $\Phi(V_0, \sigma)$ 
with respect to $\sigma$ implies
\beas
0 
&=& 
\left.\frac{d \Phi(V_0, \sigma)}
{d \sigma}\right|_{\sigma_0} =
\Tr\left[f (-\Delta + V_0 + \sigma_0)\right] - \Lambda\\
&=&
{\sum}_k f(\mu_{0,k} + \sigma_0) - \Lambda
= {\sum}_k \lambda_{0,k} - \Lambda
\eeas
so that $\sum_k \lambda_{0,k} = \Lambda$ as claimed.
In order that $(\psi_0,\lambda_0,\mu_0,V_0)$ is a 
steady state we need to show that
\be \label{poissonphi}
\lap V_0 + {\sum}_k \lambda_{0,k} |\psi_{0,k}|^2=0 .
\ee
To verify this we observe that $V_0$, being a maximizer of
$\Phi(\cdot,\sigma_0)$, satisfies the Euler-Lagrange equation
\be \label{phieq}
\lap V_0 (x) + K_{f(-\lap + V_0 + \sigma_0)}(x,x)=0,
\ x \in \Omega,
\ee
where $K_L$ is the kernel associated with a trace-class
 operator $L$. In our case
\be \label{ourkernel}
K_{f(-\Delta + V_0 + \sigma_0)}(x,x)= 
{\sum}_k f(\mu_{0,k} + \sigma_0)
|\psi_{0,k}|^2(x)
\ee
and (\ref{poissonphi}) follows from (\ref{phieq}), 
(\ref{ourkernel}), and
the fact that by definition, 
$\lambda_{0,k}=f(\mu_{0,k} + \sigma_0).$
As to the proof for $(\psi_0,\lambda_0)\in \S$
we refer to  Remark~\ref{ststins}. 
\prfe

\noindent
In view of the relations between our various functionals
derived in the previous section it is of interest to note:

\begin{remark}
{\rm
If $(V_0,\sigma_0)$ is the maximizer obtained
in Theorem~\ref{exstst} and $(\psi_0,\lambda_0,\mu_0,V_0)$ 
is the 
corresponding steady state, then
\[
\Phi(V_0, \sigma_0) = \H_C (\psi_0, \lambda_0).
\]
To see this, note that by (\ref{phidef}) we have
\[
\Phi\left(V_0, \sigma_0\right)= 
\G (\psi_0, \lambda_0, V_0, \sigma_0)
\leq \H_C (\psi_0, \lambda_0),
\]
where equality holds iff 
$V_0$ is the maximizer of 
$\G(\psi_0,\lambda_0, V, \sigma_0)$ on $ H_0^1$;
note that here $\G$ is independent of $\sigma$ since 
$\sum_k\lambda_{0,k}=1$. 
This, on the other hand,
is equivalent to the fact that $V_0$ is the solution 
of the Poisson equation (\ref{poissonphi}).
}
\end{remark}

\end{document}